# Observation of Fine Structure in Channeling of Particles in Bent Crystals

A. Mazzolari[1,2], H. Backe[3], L. Bandiera[2], N. Canale[2], D. De Salvador[4,5], P. Drexler[3], V. Guidi[1,2], P. Klag[3], W. Lauth[3], L. Malagutti[1,2], R. Negrello[1,2], G. Paternò[1,2], M. Romagnoni[1,2], F. Sgarbossa[4,5], A. Sytov[2], V. Tikhomirov, D. Valzani[4,5]

[1]*Department of Physics and Earth Sciences, University of Ferrara, Via Saragat 1/c, 44122 Ferrara, Italy*
[2]*INFN, Section of Ferrara, Via Saragat 1, 44122 Ferrara, Italy*
[3]*Institut für Kernphysik der Universität Mainz, Mainz, 55099, Germany*
[4] *Department of Physics, University of Padua, Via Marzolo 8, Padua, 35131, Italy*
[5] *INFN Laboratori Nazionali di Legnaro, Viale dell'Università 2, Legnaro, 35020, Italy*

We observed highly efficient manipulation of the 530 MeV positron beam at Mainz Microtron via bent crystals. The low beam divergence revealed a fine structure in the angular distribution of channeled particles. A compact analytical model, supported by Monte Carlo including multiple scattering, accounts for the measurements. We established a criterion for designing a crystal with reduced angular spread of channeled particles for beam manipulation at any energy. This finding is particularly useful in view of applications of channeling to the highest energies.

As a charged particle enters a crystal at an angle smaller than the critical angle for channeling, $\theta_c$, it becomes captured by the energy potential well built up by coherent interactions with atoms aligned along atomic planes or axes [1] (see Fig. 1c). This phenomenon is relevant for several applications, including microelectronics [2], recent innovative ideas in space exploration [3], nuclear research [4], and particle physics [5-9]. A particularly intriguing aspect is channeling in bent crystals, which enables steering of particle beams without relying on intense magnetic fields. As represented in Fig. 1a, positively charged particles, channeled in a bent crystal, follow its curvature while traveling through it with minimal interaction. First proposed by Tsyganov in 1976, this concept has evolved into an advanced technology that utilizes planar [10] and axial channelings [11-13], as well as volume reflection (VR) [14-16], to steer particle beams throughout an extraordinarily wide energy range, spanning from MeV [17] to several TeV [18-21]. Remarkable application has been for the Large Hadron Collider (LHC), where bent crystals empower beam collimation—a demonstration of their potential for high-energy physics [5-8, 18]. Driven by coherent interactions, channeled positive particles oscillate between atomic planes (see Fig. 1b). However, incoherent scattering with individual nuclei may cause dechanneling, namely the kick out of particles from the channel state. As a bent crystal is aligned to match the beam trajectory, particles may experience VR at the tangency point with atomic planes, deflecting them away from the bend, or being captured for channeling via volume capture (VC).

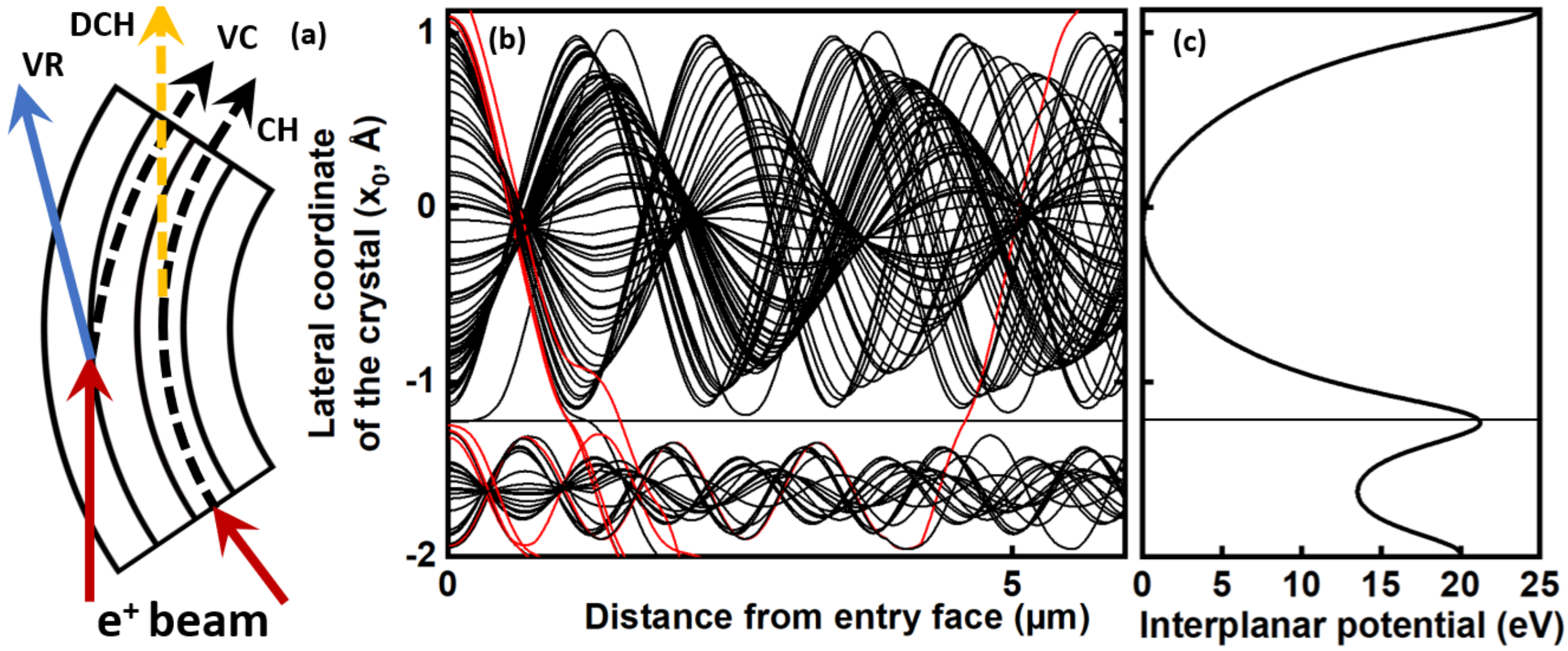

Figure 1. (a) Schematic of beam-to-bent-crystal interactions. Tangential entrance to the planes yields channeling (CH); incoherent scattering induces dechanneling (DCH); tangency inside the bulk leads to volume reflection (VR) or volume capture (VC). Interplanar spacing is assumed uniform for schematic clarity. (b) Simulated positron trajectories in Si(111): channeled (black) and dechanneled (red), illustrating the oscillatory motion under CH. (c) Interplanar potential for Si(111) with bending radius *R*=30 mm, exhibiting the characteristic double-well structure of the (111) orientation.

The undulation of particles within a channel, known as "planar channeling oscillations" [2, 22] extends beyond the initial understanding of the phenomenon. The study of planar channeling oscillations is relevant for low-energy surface spectroscopy such as Rutherford backscattering, inasmuch as they qualify the crystalline quality [2]. Indeed, planar oscillations play marginal role for particle steering of medium and high-energy beams, despite some theoretical advancements and the potential applications suggested by simulations in [23].

In this letter we demonstrate very efficient steering of a 530 MeV positron beam through channeling and volume reflection in a bent crystal. We show that channeling oscillations imprint complex dynamics on channeled particles, producing a multi-peak fine structure in the angular distribution of the beam exiting the crystal. We investigate this effect with a compact analytical model and Monte Carlo simulations [24].

A positron beam was delivered by the new beamline at the MAMI accelerator (Mainz, Germany) [25]. The beam, 0.25x1.5mm$^2$ wide and 530±10 MeV in energy, features an extra-low divergence of ~60 μrad, well below the critical angle for channeling (~300 μrad). For channeling studies we used (111) oriented crystal 29.9±0.1 μm thick exploiting quasi-mosaic effect [26-30] under regime of large deformations, allowing complete suppression of anticlastic deformation [[27]]. For this crystallographic orientation, the interplanar potential exhibits two independent potential wells [31] (Fig. 1). Positrons were detected 8 m downstream the crystal by an octagonal scintillation counter [32] and a silicon pixel detector with 80x80 μm$^2$ pixel size [33]. Fig. 2(a) shows experimental data on deflection angle, $\Delta\theta$, vs. the angle between atomic planes and beam direction at the entry face of the crystal, $\alpha$. Figure 2(b) compares beam profiles under channeling or VR against Geant4 simulations [24, 34]). Channeling peaks at 970±10 μrad, featuring 72±1% efficiency vs. 73.0±0.5% by simulations. The tail on the left owes to over-barrier particles at the crystal entry face [35]. Nuclear dechanneling length [36-38] was measured to be 18.5±1.5 μm vs. 18.8±0.2 μm from simulations. The black curve is for deflection amid the VR region. Here, VR efficiency was 78±1% (79±0.5% from simulations). Although sub-GeV positron channeling had already been observed [39], the efficiency in that study was limited to a few percent, which is insufficient for practical use. Within the "channeling peak" (area 2 in Fig. 2a), notable deviation from the usual oblate distribution observed in similar experiments was detected, revealing an unexpected “fine structure”. As an example, Fig. 2d shows experimental angular distribution at $\alpha$=160 μrad. Indeed, previous determinations were unable to observe such effect (see [40] for a review) because of too large beam divergence. The experiment in this paper was run with extraordinarily high beam quality, which played key role in observing the fine structure.

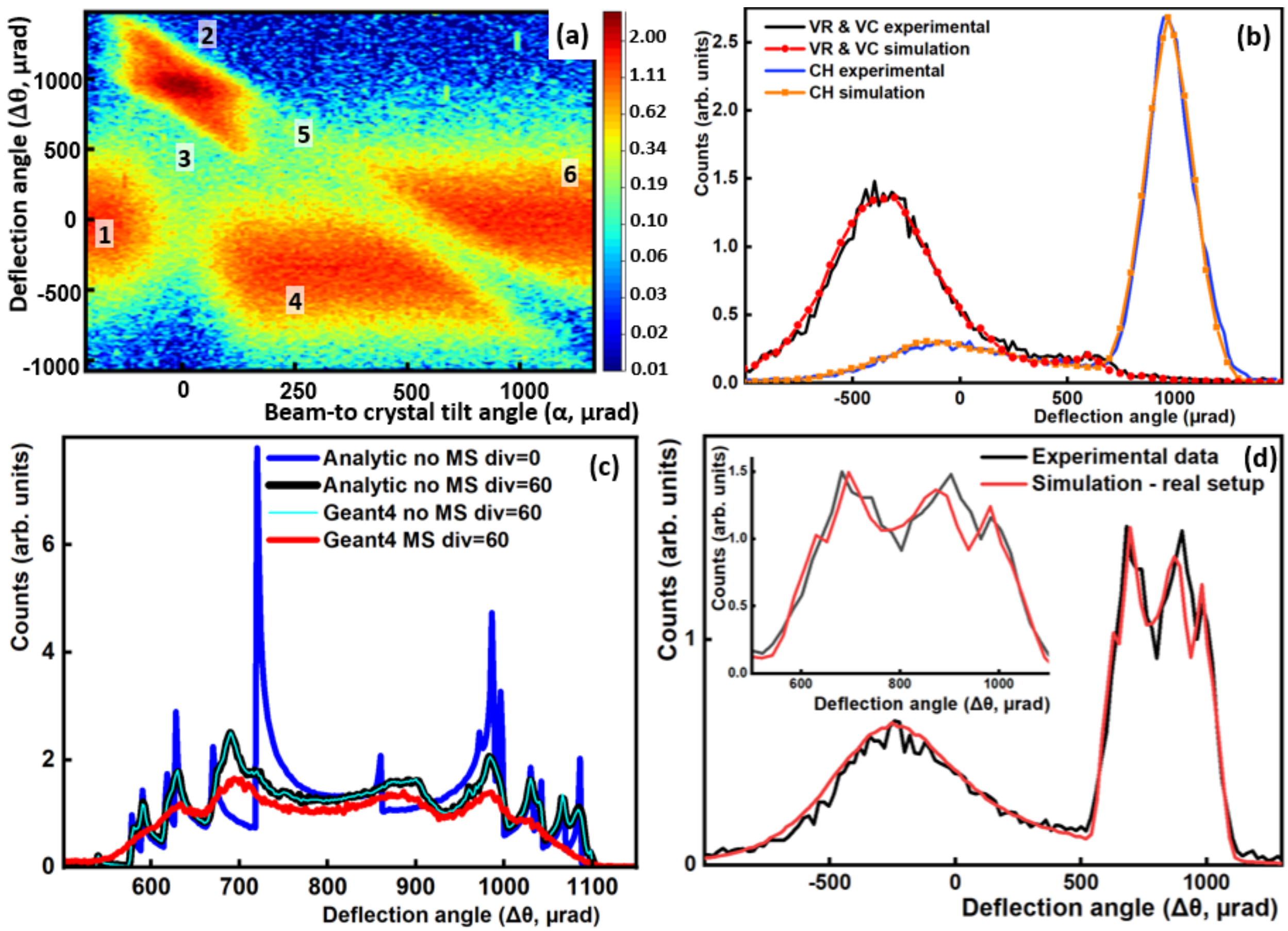

Figure 2. (a) Experimental "angular scan" of positron beam interaction with the crystal, showing six regions: (1) and (6) beam-to-crystal misalignment, (2) channeling, (3) dechanneling, (4) volume reflection, (5) volume capture. (b) Angular distributions with the crystal aligned for channeling or volume reflection. Continuous curves are experimental data, dotted lines are for simulations. (c) Fine structure of the channeling at $\alpha$=160 μrad as predicted by the model for 0 (blue curve) and for 60 μrad (black) beam divergence. Geant4 simulations without (cyan) and with (red) multiple scattering (MS) for a 60 μrad divergence beam. (d) Angular distribution at α = 160 μrad. The channeling fine structure produces three peaks, observed experimentally and reproduced by Geant4. The red curve in (d) is identical to that in (c); for direct comparison, the simulated distribution has been rebinned to match the data binning. Inset: zoom of the channeling peak highlighting its fine structure.

Preliminary analysis about the formation of fine structure appeared in [41]. The origin of such pattern has been explained analytically as follows (see End Matter for details). At first, we neglected the role of multiple scattering. We recall that the particles channeled between atomic planes oscillate with a spatial period, $\lambda = \frac{2\pi}{k}$ , and amplitude, $A$, both depending on impact parameter, $x_0$, and on beam-to-crystal tilt angle, $\alpha$ [42]. At the crystal exit, the deflection angle holds

$$\Delta\theta = \theta_b - \alpha + \left.\frac{dx}{dz}\right|_{z=t} \quad (1)$$

where $t$ is crystal thickness and $\left.\frac{dx}{dz}\right|_{z=t}$ the slope of particle trajectory with respect to atomic planes at the exit face (see Fig. 5). In our model, the motion was simplified as harmonic. Any departure from harmonicity was encoded solely in $\lambda(x_0, \alpha)$ and $A(x_0, \alpha)$. It holds $x(z) = A\sin\big(k(z + z_0)\big)$ and $\left.\frac{dx}{dz}\right|_{z=t} = Ak\cos\big(k(t + z_0)\big)$. From Eq. 1, we derive $\Delta\theta = \theta_b - \alpha + Ak\cos(k(t + z_0))$. Since $k = k(x_0, \alpha)$ and $z_0 = z_0(x_0, \alpha)$, $\Delta\theta$ becomes a function of $x_0$ and $\alpha$, too (see Figs. 3(a-b)). While $\alpha$ is directly controlled in the experiment, $x_0$ is a uniformly distributed random variable within $\left[-\frac{d}{2}, \frac{d}{2}\right]$ with probability density $p(x_0) = \frac{1}{d}$, $d$ represents the width of the potential well available for channeling. We will interpret the emerging fine structure through the formalism of Thom–Arnol'd [43, 44]. Thus, we regard $x_0$ as a state variable and $\alpha$ as a control parameter governing the mapping $(x_0, \alpha) \to \Delta\theta$ (see Fig. 3(a-b)). This geometric viewpoint motivates the analysis that follows. Setting the value of $\alpha$, the deflection-angle distribution is obtained by projecting the curve $(x_0, \Delta\theta(x_0))$ onto the $\Delta\theta$ axis (see distributions on right-hand panel in Fig. 3(a)). By conservation of probability, $p(x_0)dx_0 = p(\Delta\theta)d\Delta\theta$, we have $p(\Delta\theta) = \frac{p(x_0)}{\left|\frac{d\Delta\theta}{dx_0}\right|}$ [45], $p(\Delta\theta)$ being the probability density of $\Delta\theta$. From this standpoint, sharp peaks in $p(\Delta\theta)$ arise at stationary points of $\Delta\theta(x_0)$, i.e. where the derivative $\frac{d\Delta\theta}{dx_0} = 0$. Such peaks occur at deflection angles $\Delta\theta_{peak}$ under the condition (see End Matter for detailed calculation):

$$\tan\big(k(t + z_0)\big) = \frac{kx_0(k + x_0k')}{(Ak)^2tk' + \alpha(k + x_0k')} \quad (2)$$

where $k' = \partial k/\partial x_0$.

The most relevant family of solutions occurs at $x_0 = 0$, which implies $z_0 = m\pi/k$, with $m$ an integer number, through Eq. 6 (see End Matter). It holds $kt = n\pi$, namely the trajectories of the particles forming the peak enter and exit the crystal on the center of the potential well, i.e. $x(0) = x(t) = 0$. In this case, the deflection is

$$\Delta\theta_{peak} = \begin{cases} \theta_b & n\ even \\ \theta_b - 2\alpha & n\ odd \end{cases} \quad (3)$$

Here, the trajectories in the crystal bounce an integer number of half-oscillations ($n\lambda/2$, equivalently $k(0, \alpha)t = n\pi$).

A second notable family of solutions is for $k + x_0k' = 0$, here the dependence of $k$ on $x_0$ exactly compensates for the difference in the initial phase, leading to a peak in the outcoming beam. It leads to $k(t + z_0) = n\pi$ (see Fig. 3a), namely the peak exits the crystal with deflection $\Delta\theta_{peak} = \theta_b - \alpha + (-1)^n Ak$ at $x(t) = 0$, i.e. on the center of the potential well. As for the first solution, the particles must exit the crystal on the bottom of the potential well ($x(t) = 0$), thought they are not required to start there.
However, the two peak families exhibit different intensities. For the first family ($x_0 = 0, kt = n\pi$) , analysis of the map $x_0 \rightarrow \Delta\theta$ shows that the peaks correspond to a swallowtail-type singularity, as named in [44]. Hence $p(\Delta\theta) \propto |\Delta\theta - \Delta\theta(x_0^*)|^{-\frac{3}{4}}$ with peak height scaling as $\left|\Delta\theta^{(4)}(0)\right|^{-\frac{1}{4}}$ [44]. It also follows that the first family constraints $\alpha$ to a discrete set $\{\alpha_n\}$, satisfying $k(0, \alpha_n)t = n\pi$ (see Fig. 3(b)). For the second family, ($k + x_0 k' = 0$), the peaks correspond to regular stationary point ($\Delta\theta'(x_0^*) = 0$ and $\Delta\theta''(x_0^*) \neq 0$, i.e. an ordinary fold, as named in [44]). Hence $p(\Delta\theta) \propto |\Delta\theta - \Delta\theta(x_0^*)|^{-\frac{1}{2}}$ and peak height scales as $|\Delta\theta''(x_0^*)|^{-\frac{1}{2}}$ [44]. In summary, strong peaks are formed because of the first solution and weak peaks are formed owning to the second one (-3/4 vs. -1/2 power law intensity).
In any case, it is remarkable to observe that the imprint of the initial conditions propagates deep into the crystal and shapes the profile of the exit-angle distribution.
For comparison with experimental data, we applied the analytical model to two cases—an idealized zero-divergence beam and a beam with divergence as for the experiment (Fig. 2c blue and black curves respectively). To move beyond a purely analytical description, we performed Monte Carlo simulations based on Geant4 in two configurations: (i) a case that includes beam divergence while disabling multiple scattering (Fig. 2c, cyan curve) and (ii) a fully realistic case that includes both beam divergence and multiple scattering (Fig. 2c, red curve). In both the zero-divergence limit and in the finite-divergence case (60 μrad), the analytical model and Geant4 without multiple scattering are in excellent agreement; the zero-divergence Geant4 curve is not displayed to keep the figure uncluttered. Fig. 2d shows very good agreement between experimental data and the results of Geant4 simulations when all beam characteristics are accounted for.
The mechanism underlying the fine structure of channeling peak is analyzed through the model in Fig. 3a–b for a zero-divergence beam. Fig. 3a displays the deflection angle as a function of the impact parameter for the tilt angle of Fig 2c–d. The patterns consist of several peaks, indexed by the model (only the principal ones are labeled). Fig. 3b shows the deflection as a function of the tilt angle. The brightest peaks belong to the first family ($x_0 = 0, kt = n\pi$) and are connected by luminous ridge lines—the generic solutions of Eq. (2)—which host cusp-type singularities, for which the density scales as $p(\Delta\theta) \propto |\Delta\theta - \Delta\theta(x_0^*)|^{-\frac{2}{3}}$ with a prefactor $\propto |\Delta\theta'''(x_0^*)|^{-\frac{1}{3}}$. Fainter lines with slope $-\alpha$ correspond to the second family (characterized by $k + x_0 k' = 0$). As the beam divergence has been considered in the model (Fig. 3c), the structure of peaks smears out. Identical pattern (not shown) can be achieved through Geant4 by disabling multiple scattering. Finally, enabling multiple scattering in Geant4 (Fig. 3d) yields a pattern closely matching the experimental observation.
Notably, the fine structure originates from the oscillatory motion of channeled particles and potential anharmonicity, and appears in both bent ($\theta_b \neq 0$) and straight ($\theta_b = 0$) crystals.

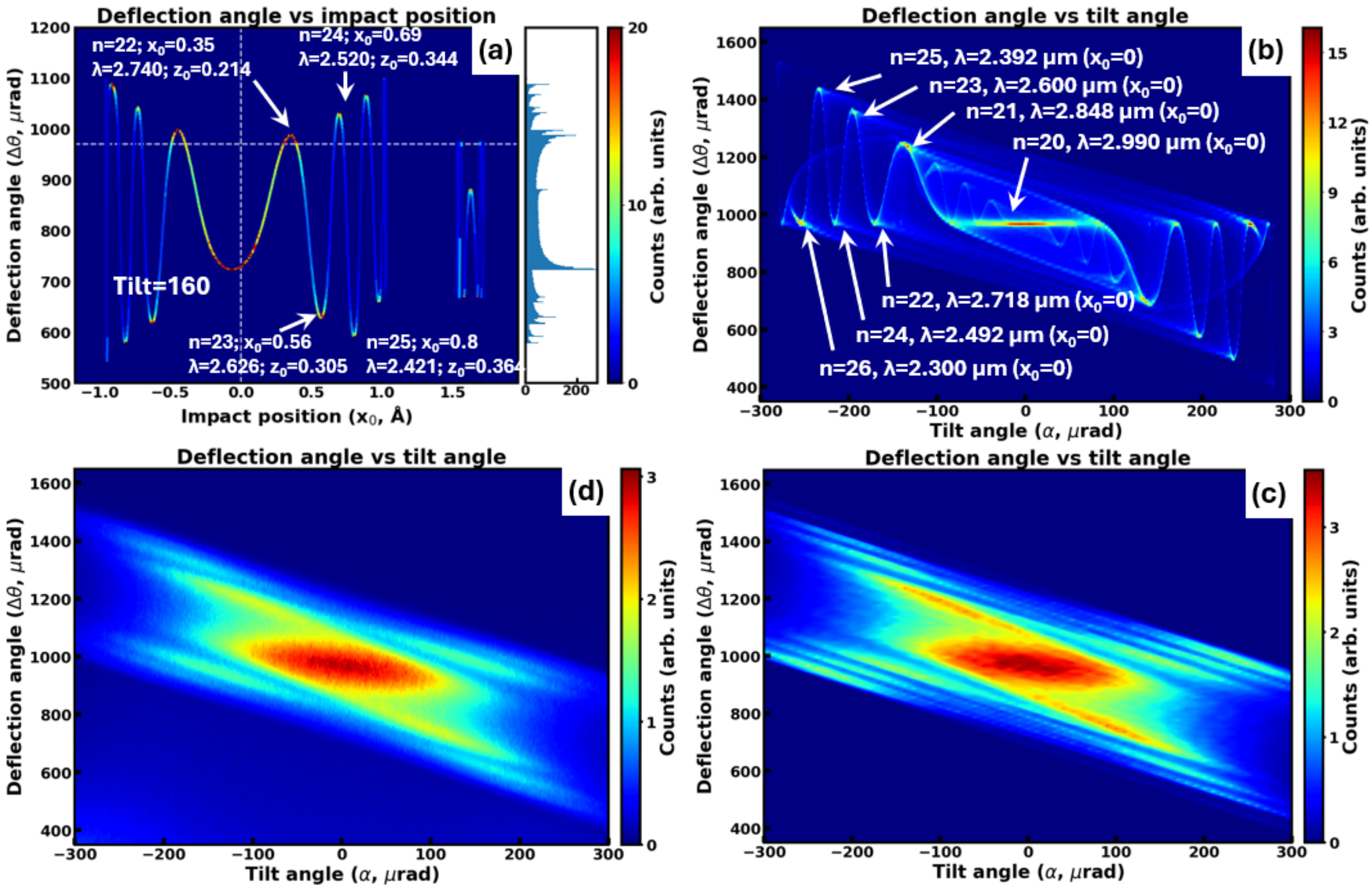

Figure 3. (a) Deflection angle $\Delta\theta$ versus impact position $x_0$ for tilt α=160 μrad. Color encodes deflection probability. The vertical dashed line marks $x_0 = 0$, (bottom of the interplanar well), the horizontal dashed line marks the bending angle. Color scale encodes $p(\Delta\theta, x_0)$. Maxima appears when the exit phase satisfies $k(t + z_0) = n\lambda$. The peaks on the right have been indexed. Marginal density $p(\Delta\theta)$, obtained by integrating over $x_0$, is shown on the right side. (b) Analytically calculated $\Delta\theta$ vs $\alpha$ for a zero-divergence beam. Intense peaks (labeled in the figure) are obtained when, for particles entering at $x_0 = 0$, the oscillation period fulfills the condition $t = \lambda(0, \alpha)n/2$. We remember that crystal thickness was $t = 29.9$ μm. The peaks with *n* even are characterized by $\Delta\theta = \theta_b$, the ones with *n* odd by $\Delta\theta = \theta_b - 2\alpha$. In both panels (a) and (b), $\lambda$ and $z_0$ are expressed in μm, $x_0$ is expressed in Å. (c) Same as in panel b, for a beam of 60 μrad divergence (experimental value). Geant4 simulations without multiple scattering (not shown) generate the same figure. (d) Monte Carlo simulation performed with Geant4, including beam divergence and multiple scattering.

Based on above considerations, we established a design rule to fabricate crystals to achieve the narrowest distribution for channeled particles. The model was carried out in terms of planar channeling oscillations, whose spatial period, $\lambda$, can be applied to any energy range for which channeling does work. For the sake of comparison bent crystals [46] already under usage at LHC to steer 7 TeV protons and heavy ions [18, 47] exhibit a thickness of $t{\sim}15\lambda$, comparable to the one in this experiment ($t{\sim}10\lambda$). As an example, Fig. 4 compares the cases of a crystal in which an integer number of half-channeling oscillations are performed ($n = 20$, red curve) and a situation intermediate between two bright peaks ($n = 20.5$, black curve). In the latter case the peak is broader.

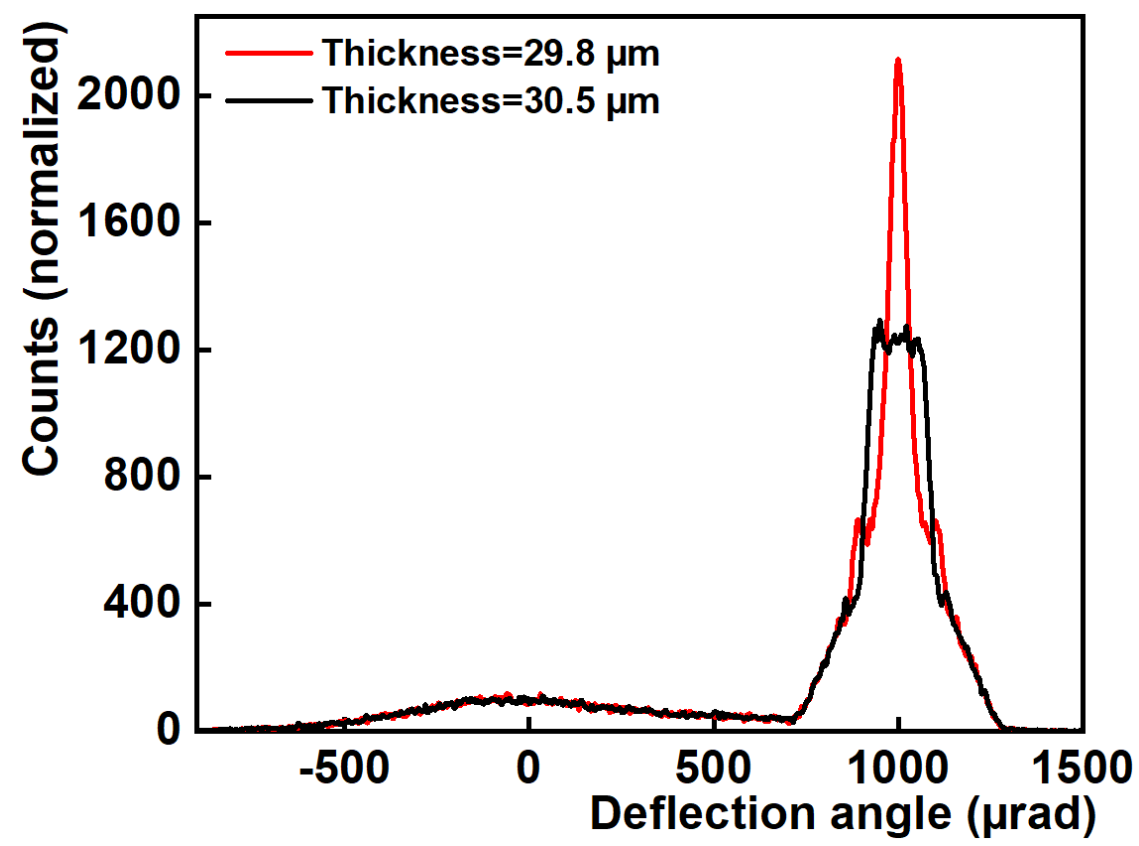

Figure 4: simulated angular distribution of channeled particles in a 29.8 μm ($t{\sim}10\lambda_0, n = 20$) and 30.5 μm ($t{\sim}10.25\lambda_0, n = 20.5$) thick crystal, both has bending angle of 970 μrad. Thickness modulates the shape of the channeling peak.

This observation is directly relevant for bent-crystal charged-particle optics requiring precise trajectory control. It also enables the design of next-generation radiation sources [48-50]. In the beam-optics domain, examples include slow extraction from particle accelerators [51, 52], manipulation of high energy particle beams [18, 21, 53] or recently proposed schemes with consecutive crystals for steering ultra-high energy particle beams [5-9, 54], where channeling efficiency in the second crystal depends on the angular distribution of the beam generated in the first crystal through channeling. In all such applications, crystal fabrication with the needed precision can be afforded because crystal thickness can be controlled with micron accuracy, a value well below fractions of $\lambda$ at the corresponding energy of operation.

**Acknowledgements**

This work was partially supported by the European Commission through the Horizon Europe EIC-Pathfinder TECHNO-CLS (G.A. 1016458) and H2020 MSCA RISE N-LIGHT (G.A. 872196) and H2020-MSCA-IF TRILLION (G.A. 101032975) projects. We acknowledge partial support of INFN-CSN5 through the OREO and GEANT4-INFN projects. We acknowledge financial support under the National Recovery and Resilience Plan (NRRP), Call for tender No. 104 published on 02.02.2022 by the Italian Ministry of University and Research (MUR), funded by the European Union – NextGenerationEU – Project Title: "Intense positron source Based On Oriented crySTals - e+BOOST" 2022Y87K7X – CUP I53D23001510006. We are grateful for the fruitful discussions with K. Aulenbacher, Ph. Heil and B. Ledroit and their support in the early stage of setting up the positron beam line. A. Mazzolari acknowledge Andrea Persiani and Claudio Manfredi from PERMAN S.R.L. (Loiano, Italy) for manufacturing the crystal bender used in this experiment, Gessica Gomez, Simmie Proctor and JoAnn Martin (Rogue Valley Microdevices, Medford, USA) for support with manufacturing of the crystals. G. Klug, J. Griesmann and S. Cali from DISCO GMBH (Munich, Disco) for support with crystal dicing.

# End Matter

As a charged particle of energy $E$ enters a crystal at an angle $\alpha \leq \theta_c$ with respect to its atomic planes ($\theta_c$ is the critical angle for planar channeling), it is subjected to the planar potential $U(x)$ [1]. In our case we analytically calculated $U(x)$ as Doyle-Turner [55].

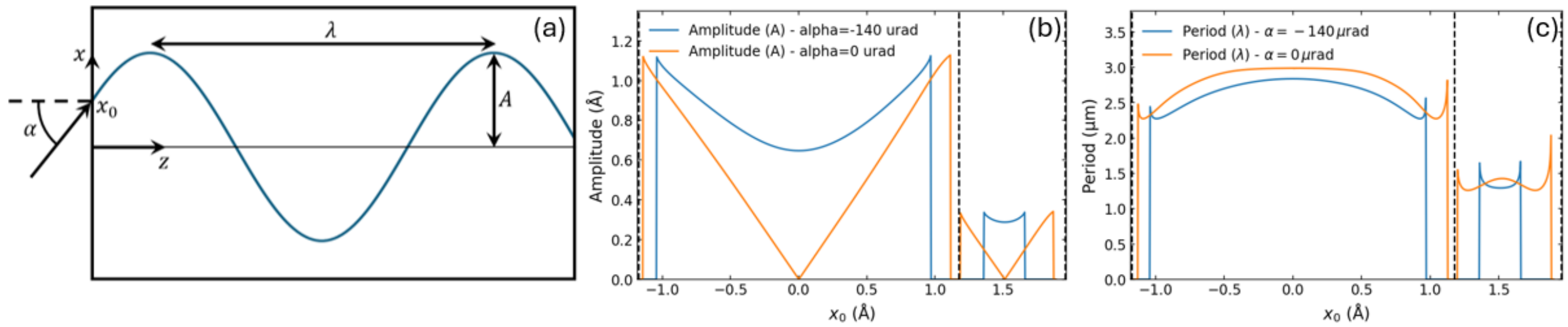

Figure 5. (a) Schematic of the transverse oscillatory motion of channeled particles between two neighboring planes. Oscillation amplitude $A$ (b) and period $\lambda$ (c) as functions of the impact position $x_0$ for two tilt angles, $\alpha = -140$ μrad (blue) and $\alpha = 0$ μrad (orange). The vertical dashed lines indicate the positions of the atomic planes.

Each trajectory is characterized by the initial conditions

$$x(0) = x_0 = A\sin(kz_0) \text{ and } \alpha = \left.\frac{dx}{dz}\right|_{z=0} = Ak\cos(kz_0) \qquad (4)$$

Transverse energy of motion is given by $E_\perp = E\sin^2(\alpha) + U(x_0)$. The trajectory turning points are given by $U(x_\pm) = E_\perp$ and the amplitude of motion is given by $A = A(\alpha, x_0) = \frac{x_+ - x_-}{2}$. Spatial period is [42]:

$$\lambda = \lambda(\alpha, x_0) = 4\int_0^A \sqrt{\frac{E}{E\alpha^2 + U(x_0) - U(x)}}\,dx \qquad (5)$$

We assume the trajectory to be sinusoidal, i.e. $x(z) = A\sin\big(k(z + z_0)\big)$, where $k = 2\pi/\lambda$. Eq. 4 leads to:

$$\sqrt{\alpha^2 + (kx_0)^2} = Ak, \Delta\theta = \theta_b - \alpha + Ak\cos(k(t + z_0)), \tan(kz_0) = \frac{kx_0}{\alpha} \quad (6)$$

The deflection angle given by the crystal holds $\Delta\theta = \theta_b - \alpha + \left.\frac{dx}{dz}\right|_{z=t}$. While $\alpha$ is directly controlled in the experiment, $x_0$ is uniformly distributed in $\left[-\frac{d}{2}, \frac{d}{2}\right]$, with probability density $p(x_0) = \frac{1}{d}$, *d(α)* being the width of the interplanar well for channeling. By conservation of probability, $p(x_0)dx_0 = p(\Delta\theta)d\Delta\theta$, we have $p(\Delta\theta) = \frac{p(x_0)}{\left|\frac{d\Delta\theta}{dx_0}\right|}$ and peaks are formed at the stationary points of $\Delta\theta(x_0)$, i.e. where $\frac{d\Delta\theta}{dx_0} = 0$. For compactness, we use the following notations: $k' := \frac{dk}{dx_0}$, $z_0' := \frac{dz_0}{dx_0}$, $R := Ak$, and the phase $u := k(t + z_0)$. We have:

$$\frac{d\Delta\theta}{dx_0} = \frac{dR}{dx_0}\cos(u) - R\,\frac{du}{dx_0}\sin(u) \qquad (7)$$

It holds:

$$\frac{dR}{dx_0} = \frac{1}{2R}2(kx_0)(k + x_0k') = \frac{kx_0(k + x_0k')}{R}, \ \frac{du}{dx_0} = (t + z_0)k' + kz_0' = tk' + \frac{d}{dx_0}(kz_0) \qquad (8)$$

Recalling that $\tan(kz_0) = kx_0/\alpha$, we have $\frac{d}{dx_0}\tan(kz_0) = \frac{d}{dx_0}\left(\frac{kx_0}{\alpha}\right)$, $\Rightarrow \sec(kz_0)\frac{d}{dx_0}(kz_0) = \frac{k + x_0k'}{\alpha}$. Bearing in mind that $1/\sec(kz_0) = \cos^2(kz_0)$ we have:

$$\frac{d}{dx_0}(kz_0) = z_0k' + kz_0' = \frac{k+x_0k'}{\alpha}\cos^2(kz_0) = \frac{\alpha(k+x_0k')}{(Ak)^2} \quad (9)$$

This leads to: $\frac{du}{dx_0} = tk' + \frac{\alpha(k+x_0k')}{R^2}$ and:

$$\frac{d\Delta\theta}{dx_0} = \frac{kx_0(k+x_0k')}{R}\cos(u) - R\left(tk' + \frac{\alpha(k+x_0k')}{R^2}\right)\sin(u) \quad (10)$$

On demanding $\frac{d\Delta\theta}{dx_0} = 0$, it leads to the condition:

$$\tan\big(k(t+z_0)\big) = \frac{kx_0(k+x_0k')}{(Ak)^2tk' + \alpha(k+x_0k')} \quad (11)$$

A notable solution is $k(t+z_0) = n\pi$ which, using Eq. 6, implies the following cases

$$x_0 = 0 \Rightarrow kt = n\pi \Rightarrow \Delta\theta_{peak} = \begin{cases}\theta_b & n\ even\\ \theta_b - 2\alpha & n\ odd\end{cases} \quad (12.1)$$

$$(k+x_0k') = 0 \Rightarrow \Delta\theta_{peak} = \theta_b - \alpha + (-1)^n Ak \quad (12.2)$$

**Evaluation of peak intensity**

For the case described by Eq. 12.1, it holds $\left.\frac{d\Delta\theta}{dx_0}\right|_{x_0^*} = \left.\frac{d^2\Delta\theta}{dx_0{}^2}\right|_{x_0^*} = \left.\frac{d^3\Delta\theta}{dx_0{}^3}\right|_{x_0^*} = 0$ and $\left.\frac{d^4\Delta\theta}{dx_0{}^4}\right|_{x_0^*} \neq 0$, leading to:

$$p(\Delta\theta) \sim \frac{1}{\sqrt[4]{\left|\Delta\theta^{(4)}(x_0^*)\right|}}\left|\Delta\theta - \Delta\theta_{peak}\right|^{-3/4} \quad (13)$$

For the case described by Eq. 12.2, it holds $\left.\frac{d\Delta\theta}{dx_0}\right|_{x_0^*} = 0$ and $\left.\frac{d^2\Delta\theta}{dx_0{}^2}\right|_{x_0^*} \neq 0$, where $x_0^*$ is a stationary point for $\frac{d\Delta\theta}{dx_0}$, then $\Delta\theta(x_0) \simeq \Delta\theta_{peak} + \frac{1}{2}\Delta\theta''(x_0^*)(x_0 - x_0^*)^2$, leading to:

$$p(\Delta\theta) \sim \frac{1}{\sqrt{\left|\Delta\theta^{(2)}(x_0^*)\right|}}\left|\Delta\theta - \Delta\theta_{peak}\right|^{-1/2} \quad (14)$$

**Comparison between model and Geant4 simulation with and without multiple scattering.**

Figure 6 compares angular scan reporting distribution of deflection angles obtained from analytical model (Fig. 6(a)), Geant4 simulations disabling multiple scattering (Fig. 6(b)), and more realistic case of Geant 4 simulations enabling multiple scattering (Fig. 6(c)).

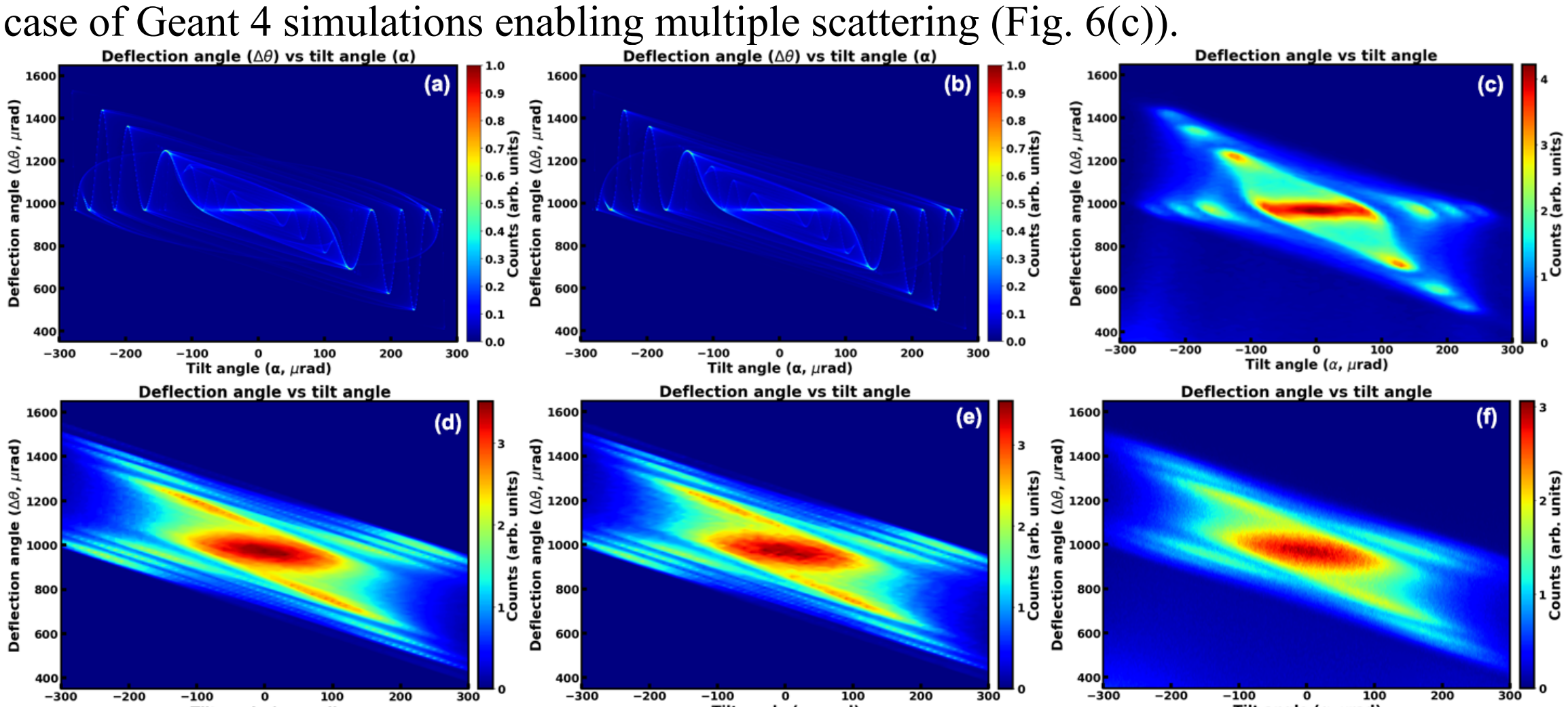

Figure. 6 "Angular scan" obtained under different conditions (a) by the analytical model for a beam of zero-divergence. (b) By Geant4 for a beam of zero-divergence and disabling multiple scattering. (c) By Geant4 for a zero-divergence beam and enabling multiple scattering. (d) By the analytical model for a beam with experimental beam divergence.

(e) By Geant4 for a beam with experimental beam divergence and disabling multiple scattering. (f) By Geant4 for a beam with experimental beam divergence and enabling multiple scattering.